\documentstyle[aps,prb,epsfig,floats,twocolumn,amsmath]{revtex}

\begin{document}

\flushbottom

\twocolumn[
\hsize\textwidth\columnwidth\hsize\csname @twocolumnfalse\endcsname

\title{
Path integral Monte Carlo simulation of helium at negative
pressures}

\author{Gregory H. Bauer, David M. Ceperley and Nigel Goldenfeld}
\address{Department of Physics, University of Illinois, 1110 W. Green
St., Urbana, Illinois 61801-3080}

\maketitle

\begin{abstract}

Path integral Monte Carlo (PIMC) simulations of liquid helium at
negative pressure have been carried out for a temperature range
from the critical temperature to below the superfluid transition.
We have calculated the temperature dependence of the spinodal line
as well as the pressure dependence of the isothermal sound
velocity in the region of the spinodal. We discuss the slope of
the superfluid transition line and the shape of the dispersion
curve at negative pressures.


\end{abstract}

\vspace{1cm}
]

\columnseprule 0pt

\narrowtext


\section{Introduction}



\begin{figure}
\begin{center}
\epsfig{file=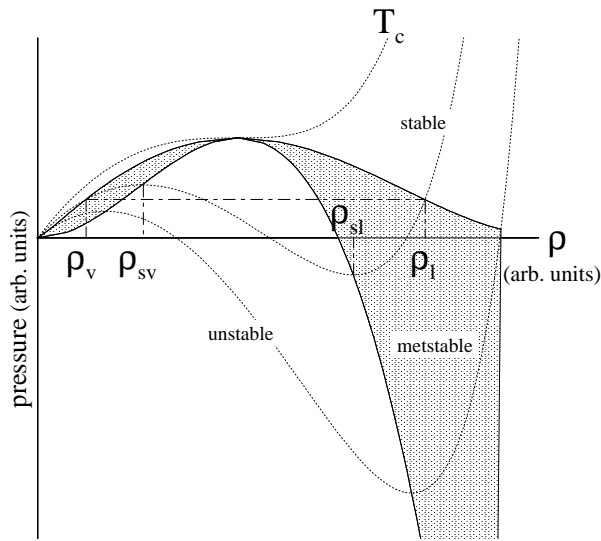,width=3.2in}
\end{center}
\begin{minipage}{3.4in}
\vspace{0.5cm}
\caption{
Isotherms of pressure versus density for a generic liquid are
shown for temperatures at or less than the critical temperature. The
spinodal line and the coexistence line form the lower and upper bounds
of the shaded metastable region. Locations of the saturated vapor and
liquid densities, $\rho_{v}$ and $\rho_{l}$, and the spinodal vapor
and liquid densities, $\rho_{sv}$ and $\rho_{sl}$, for a given
isotherm are indicated.
\label{diagram}}
\end{minipage}
\end{figure}

Consider a liquid that is quenched from above the liquid-gas
coexistence line to a point below, which will be in either the
metastable or unstable region, see Fig.~\ref{diagram}.  As it
approaches equilibrium, the system will phase separate to a state of
positive pressure with coexisting vapor and liquid phases of densities
$\rho_{v}$ and $\rho_{l}$ respectively.  Below this coexistence line
is the spinodal line, which delineates the metastable phase from the
unstable phase. The spinodal line is the locus of points where the
speed of sound vanishes, $m_{4} c^{2}$ = $\partial P / \partial \rho$
where $m_{4}$ is the mass of a $^{4}$He particle. At the spinodal
there is no energy barrier to nucleation and phase separation. If the
temperature is low enough the pressure at the liquid spinodal, at a
density of $\rho_{sl}$, may be negative. Once the spinodal pressure
becomes negative, the lowest pressure the system can attain is the
spinodal pressure.



Direct measurement of the spinodal pressure of liquid helium by
homogeneous nucleation is experimentally difficult. In a driven
system, such as the pressure oscillation experiments of
Maris\cite{Maris93,Maris96} and Balibar\cite{Balibar95,Balibar98},
negative pressures are only achieved for a finite duration. The
presence of objects like vortices\cite{Maris94a},
electrons\cite{Hall95,Classen98b} or both\cite{Classen98a} lower the
nucleation energy barrier. Measurements of the cavitation pressure are
higher than the spinodal pressure because of quantum tunneling
\cite{Lambare98} or thermal activation \cite{Pettersen94a} over the
barrier, but are consistent with predictions of the nucleation energy
barrier, attempt frequency and the spinodal pressure.  In this paper
microscopic PIMC simulations of liquid helium at negative pressure and
finite temperature will be presented.  The calculated temperature and
density dependence of the spinodal line compare favorably with other
calculations\cite{Guirao92,Hall97,Boronat94a,Boronat94b,Solis92,Dalfovo95,Campbell96}. We estimate that the superfluid transition can be
extended to negative pressures.  Finally, properties of the excitation
spectrum are found to be consistent with conjectures based on the
quasi-particle model of superfluid helium\cite{Hall97}.






\section{Method}



At zero temperature, density functional\cite{Guirao92,Dalfovo95}
and microscopic/phenomenological\cite{Solis92,Campbell96}
calculations and quantum Monte Carlo\cite{Boronat94a,Boronat94b}
simulations of liquid helium at negative pressure have produced
comparable values of the spinodal pressure and density.  At finite
temperature, a density functional\cite{Guirao92} developed for the
liquid-gas interface has been used to successfully examine helium
at negative pressures. Microscopic finite temperature simulations
at negative pressures have not been performed until now.


It has been well documented that PIMC can provide accurate
thermodynamic and some dynamical properties of quantum systems in
equilibrium, and has been especially successful with liquid helium
\cite{Ceperley95,Gordillo98}. The basis for the path integral method
is the evaluation of the many particle density matrix,
$\rho=\exp(-\beta {\mathcal H})$. The Hamiltonian is assumed to be
\begin{equation}
{\mathcal H} = -\lambda \sum_{i=1}^{N} \nabla_{i}^{2} + \sum_{i<j}
v(r_{ij}),
\end{equation}
where an accurate two-body interaction $v(r)$ \cite{Aziz95} is used.
From the density matrix, expectation values of observables can be
determined,
\begin{equation}
\langle O \rangle = Z^{-1}\int dRdR'\rho(R,R')\langle R'|O|R\rangle,
\label{Oaverage} \end{equation}
with the partition function $Z$,
\begin{equation}
Z = \int dR\rho(R,R),
\end{equation}
and where the matrix elements of the density matrix in position basis is
\begin{equation}
 \rho(R,R')=\langle R|e^{-\beta\mathcal{H}}|R'\rangle. \label{dmatrix}
\end{equation}
$R$ is the 3$N$ state of the system, $R=\{r_{1},r_{2},..r_{N}\}$.
Using the product property, the density matrix can be expressed as
\begin{equation}
 e^{-\beta\mathcal{H}}=(e^{-\tau\mathcal{H}})^{M}
\end{equation}
where $M\tau=\beta=1/k_{B}T$ and $\tau$ is the imaginary {\it time}
step. By expressing the density matrix at $T$ as a product of density
matrices at a higher temperature of $MT$, Eq.~\ref{Oaverage} and
Eq.~\ref{dmatrix} can be accurately evaluated by Monte Carlo. In our simulation, the canonical ensemble is used, with a fixed number of particles
$N$, simulation volume and temperature $T$. For $^{4}$He, $\lambda =
6.05961$ K \AA$^{2}$, and during the simulations $\tau = 0.0125$
K$^{-1}$ and $N=64$. PIMC incorporates Bose statistics,
necessary for the modeling of superfluid $^4$He, by allowing the
permutation of particle paths.  Superfluidity manifests itself as the
{\it winding}\cite{Ceperley89} of the particle paths across the
simulation cell when periodic boundaries are present.

An advantage of a finite system is its ability to explore systems at
negative pressures.  A disadvantage is that finite size effects are
present, which can be appreciable for a small system and can be
difficult to correct.  The liquid-gas interface limits how large a
system can be simulated due to the surface energy of a phase separated
system. Because the simulation is in equilibrium it will phase
separate if it is thermodynamically favorable. Choosing a system size
that is on the order of the liquid-gas interface width prevents the
system from forming a stable coexistence.  From previous PIMC
simulations \cite{Ceperley95} it was found that a 64 particle system
is large enough to provide accurate bulk properties. To check that
this holds true near the spinodal, the pressure near the spinodal was
examined for several systems from 8 to 256 particles at a temperature
of 2.0~K. In the range of 32 to 128, the pressure remained negative
and varied little with particle size, and $S(k) < 1$ for small $k$
(see below).  A system of 64 particles was found to be optimal for
minimizing finite size effects and keeping the fluid homogeneous at
negative pressure\cite{Campbell96}.

\begin{figure}
\begin{center}
\epsfig{file=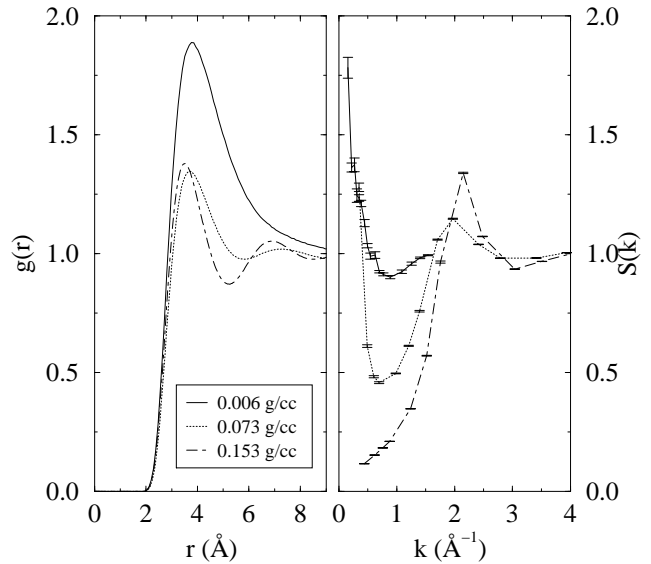,width=3.2in}
\end{center}
\begin{minipage}{3.4in}
\vspace{0.5cm}
\caption{
The pair correlation function and the static structure
function are shown for 3 densities at a temperature of 2.0~K:
0.006~g/cc, 0.073~g/cc and 0.153~g/cc.
\label{gofrsofk}}
\end{minipage}
\end{figure}

To differentiate between liquid and gas phases, the static structure
function, $S(k)$, and the pair distribution function, $g(r)$, were
examined at three representative densities.  Information about
fluctuations into the coexisting liquid-gas phase, which we wish to
avoid, is present in both $S(k)$ and $g(r)$. In Fig.~\ref{gofrsofk}
the two functions are shown for three densities and a temperature of
2.0~K. The high density liquid phase and the low density gas phase
show the typical features in $S(k)$ and $g(r)$. The intermediate
density, which is lower than the liquid spinodal density at that
temperature, begins to show signs of phase separation but no large
scale features ($S(k) \gg 1$ for small $k$) are present. As the
density is varied from liquid to gas the functions should go smoothly
from one phase to the other, for a finite system.

\section{Analysis}

\subsection{Spinodal Line}

\begin{figure}[t]
\begin{center}
\epsfig{file=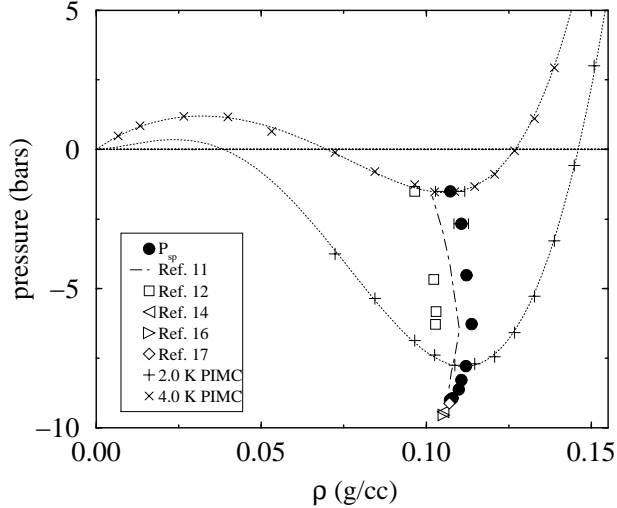,width=3.2in}
\end{center}
\begin{minipage}{3.4in}
\vspace{0.5cm}
\caption{
Spinodal pressure from PIMC simulations are compared to
other calculations and estimations of the spinodal pressure.
Increasing temperature runs from bottom to top. Two isotherms from
PIMC simulations are shown as well, with dotted lines to guide the
eye.
\label{spinodal1}}
\end{minipage}
\end{figure}

\begin{figure}[t]
\begin{center}
\epsfig{file=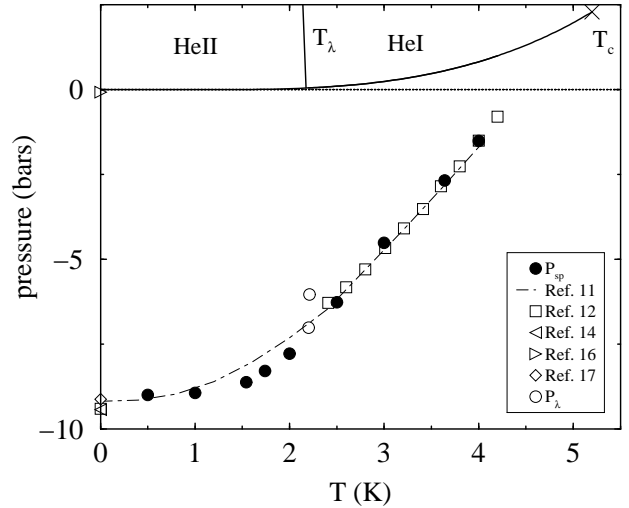,width=3.2in}
\end{center}
\begin{minipage}{3.4in}
\vspace{0.5cm}
\caption{
Temperature dependence of the spinodal pressures and the
superfluid transition at negative pressures are shown. At low
temperatures the spinodal pressure is insensitive to temperature while
at higher temperatures the behavior is linear. The upper solid lines
form usual phase diagram.
\label{spinodal2}}
\end{minipage}
\end{figure}

\begin{figure}[t]
\epsfig{file=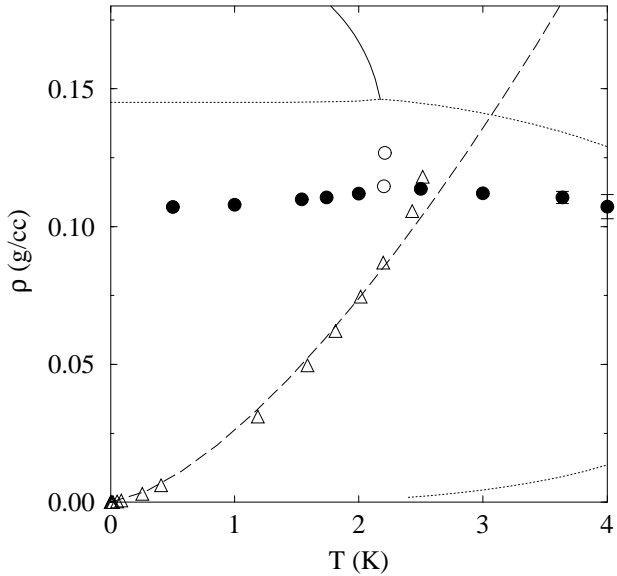,width=3.2in}
\begin{minipage}{3.4in}
\vspace{0.5cm}
\caption{
The temperature dependence of the spinodal density ($\bullet$) and
transition density ($\circ$) are shown with the experimental coexistence
curve (dashed line) and superfluid transition curve (solid line).  The
spinodal density exhibits a cusp-like feature as it intersects the
superfluid transition curve. The transition temperature for the hard
sphere Bose gas of Ref.~\protect\onlinecite{Gruter97} ($\triangle$)
and the ideal Bose gas (long dashed line) are shown for comparison.
\label{spinodal3}}
\end{minipage}
\end{figure}

\begin{table}[b]
\caption{$^{4}$He liquid spinodal values.\label{table_spinodal}}
\begin{tabular}{ccc}
$T$ ($K$) &$\rho_{sl}$ (g/cc) &$P_{sl}$ (bars) \\
\tableline
0.50& 0.1070(51)& -8.99(3) \\
1.00& 0.1077(45)& -8.93(3) \\
1.54& 0.1097(38)& -8.62(3) \\
1.74& 0.1110(51)& -8.28(4) \\
2.00& 0.1123(45)& -7.77(3) \\
2.50& 0.1136(51)& -6.27(3) \\
3.00& 0.1123(51)& -4.52(1) \\
3.64& 0.110(14)& -2.68(6) \\
4.00& 0.107(28)& -1.51(1) \\
\end{tabular}
\end{table}

The PIMC density-pressure isotherms are plotted in
Fig.~\ref{spinodal1}. To determine the location of the liquid
spinodal line, a cubic polynomial in density was fit to each
density-pressure isotherm in the region of the liquid spinodal. As
will be shown below, the cube of the isothermal speed of sound is
seen to vary linearly with pressure and from this is can be shown
that $P\sim P_{sl}+ \alpha (\rho-\rho_{sl})^{3}$ near the liquid
spinodal\cite{Hall97}.  In Table~\ref{table_spinodal} the liquid
spinodal pressure and density for various temperatures are listed
and are plotted in Fig.~\ref{spinodal1}. These points are
indicated by the filled circles. The $\chi^{2}$ goodness of fit is
on the order of 1 for low temperatures and on the order of 2 at
the highest temperatures. Simulations of the gas phase were not
done for all temperatures and so information about the gas
spinodal and the liquid-gas coexistence densities is not
available.

For comparison, density functional\cite{Guirao92,Dalfovo95},
extrapolated experimental\cite{Hall97}, optimized hypernetted
chain\cite{Campbell96} and QMC\cite{Boronat94b} values are shown
in Fig.~\ref{spinodal1}. The extrapolated values depend on the
assumption that the isothermal speed of sound has the following
power law dependence, $c^{3}\sim P-P_{sl}$, where $P_{sl}$ is the
liquid spinodal pressure. The density functional values are the
result of a parameterized density functional that reproduces the
experimental bulk properties but with the Bose condensation added
{\it post hoc}. While there is agreement with the temperature
dependence of the liquid spinodal pressure, the PIMC values of the
liquid spinodal density tend to be higher than the density
functional and extrapolated values. Linearly extrapolating the
PIMC data to 0~K yields a spinodal density very close to the QMC,
density functional and hypernetted chain values. The extrapolated
0~K PIMC spinodal pressure is slightly larger than the other 0~K
results. A possible reason for this discrepancy, at least with the
QMC value, is the use of different versions of the Aziz pair
potential\cite{Aziz95}.

In Fig.~\ref{spinodal2} the temperature dependence of the liquid
spinodal pressure, as calculated by PIMC, is plotted. For $T$ in
the neighborhood of the critical temperature $T_{c}$, the spinodal
pressure behaves similar to that of a Van der Waals gas with the
spinodal pressure decreasing monotonically with decreasing
temperature. In this region we see good agreement of the spinodal
pressures. At temperatures near zero, the spinodal pressure is
approximately independent of temperature, being slightly higher
than other values when extrapolated to 0~K . This flat region may
be understood in terms of the quasi-particle picture of liquid
helium \cite{Hall97} and will be discussed below. At temperatures
near 2~K there is a distinct transition between the two regions.
In comparison, the density functional values of
Ref.~\onlinecite{Guirao92} of the spinodal pressure vary smoothly
from the zero temperature value to the classical gas behavior,
having no flat region at low temperatures. The extrapolated
experimental spinodal pressures of Hall and Maris \cite{Hall97}
were limited to temperatures above 2.2~K or at 0~K.

It is illustrative to plot the liquid spinodal density versus
temperature, as is shown in Fig.~\ref{spinodal3}. The liquid
spinodal density shows a temperature dependence similar to the
experimental liquid coexistence density (dashed line): a
temperature independent region at low temperatures with a slight
{\it cusp} like behavior near where the experimental transition
line (solid line) approaches. The transition line at negative
pressures as determined from PIMC simulation is discussed below.


\subsection{Superfluid Transition}

It is not clear what is the temperature dependence of the
superfluid transition line at negative pressures
\cite{Campbell96,Hall97}. Experimentally, at positive pressures,
as the density is reduced, the transition temperature increases.
Further reducing the density, the transition line could extend
into the negative pressure region. Additionally, along an
isotherm, as the density is reduced to the spinodal density,
superfluidity will vanish as the speed of sound
vanishes\cite{Campbell96}. In order to accurately determine the
superfluid transition temperature we used finite size scaling of
the superfluid fraction\cite{Pollock92}. The direct estimator for
the superfluid fraction uses the {\it winding number}, which is a
measure of the degree to which the particle exchange cycles form
paths that span the width of the periodic cell \cite{Pollock87}.
For a finite $N$, the superfluid fraction should scale as,
\begin{equation}
\frac{\rho_{s}(T)}{\rho} \sim L^{-1}Q(L^{1/\nu} t) \label{scaling1}
\end{equation}
where $t=(T-T_{\lambda})/T_{\lambda}$ is the reduced
temperature. For the critical exponent of the correlation length,
$\nu$, experiment gives $\nu=0.67$.  By assuming that the scaling
function $Q$ is linear with respect to its argument near
$T=T_{\lambda}$, the parameters $\nu$ and $T_{\lambda}$ can be varied
to minimize the distance of the data to $Q$. The universal constant
$\Tilde{Q}(0)$, see Table~~\ref{table_lambda}, has been calculated for
the 3D XY model \cite{Cha91} as 0.49(1).

Using systems of 16, 32 and 64 particles at a fixed density, the
superfluid fraction was determined at several temperatures and
scaled according to Eq.~\ref{scaling1}.  Once the transition
temperature was found, the corresponding pressure was found by
interpolating the $N=64$ pressure data.  The superfluid transition
values plotted in Fig.~\ref{spinodal2} and Fig.~\ref{spinodal3}
are listed in Table~\ref{table_lambda}.  The superfluid transition
data fall along the experimental superfluid transition line if it
extended linearly to negative pressures. It appears that the
transition line intersects the spinodal line at a temperature of
2.2~K.

For comparison, the transition temperature for a hard sphere Bose
gas\cite{Gruter97} (a system which does not have a liquid-gas
transition) and an ideal Bose gas are shown in
Fig.~\ref{spinodal3}. As was explained by Gr\"{u}ter\cite{Gruter97} in
a study of superfluidity for a hard sphere Bose gas , at low densities
spatial fluctuations are important and clusters are likely to form,
which inhibit macroscopic exchange cycles. At moderate densities the
system is more homogeneous, allowing the exchange cycles necessary for
superfluidity to form and the transition temperature becomes greater
than the ideal Bose gas value by 7~\%. At high densities, exchange is
once again inhibited due to an increase in effective mass of the
particles lowering the transition temperature below the ideal Bose gas
value. Note that in comparing the temperature dependence of the
transition densities, the pressures of the ideal and hard sphere Bose
gases are positive while the pressure of the $^{4}$He transition goes
from positive to negative as the density decreases.



\begin{table}[b]
\caption{Superfluid transition values.\label{table_lambda}}
\begin{tabular}{ccccc}
$T_{\lambda}$ ($K$) &$\rho_{\lambda}$ (g/cc)
&$P_{\lambda}$ (bars)
&$\nu$
&$\Tilde{Q}(0)$\tablenote{$\Tilde{Q}(0)=\hbar^{2}\rho_{\lambda}^{2/3} Q(0)/m_{4}
 k_{B}T_{\lambda}$} \\
\tableline
2.20(2)& 0.11463& -7.014(9)& 0.63(4)& 0.49(2)\\
2.21(2)& 0.12669& -6.034(11)& 0.65(4)& 0.55(2)\\
\end{tabular}
\end{table}


\subsection{Isothermal Sound Velocity}

\begin{figure}
\begin{center}
\epsfig{file=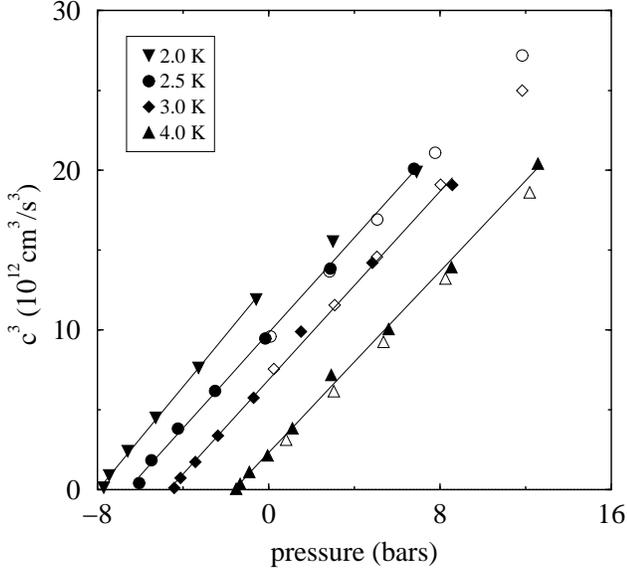,width=3.2in}
\end{center}
\begin{minipage}{3.4in}
\vspace{0.5cm}
\caption{
The isothermal speed of sound from Eq.~\ref{speed} is plotted
versus pressure for varying temperature. It has been shown
\protect\cite{Maris91a} that $c^3$ should vary linearly with pressure,
vanishing at the spinodal pressure. The straight lines are fits to the
PIMC data (filled symbols) and are to guide the eye. The experimental
data collected in Ref.~\protect\onlinecite{Hall97} are indicated by
the open symbols and are limited to positive pressures.
\label{c3}}
\end{minipage}
\end{figure}

The scaling of experimental isothermal sound velocity data to pressure
is controversial. Early thermodynamic arguments put forth by Maris
\cite{Maris89} indicated that at $T=0$ K the speed of sound should
scale with pressure as $c^{4}\sim P-P_{sl}$ where $P_{sl}$ is the
liquid spinodal pressure.  Maris' later analysis
\cite{Maris91a,Maris91b} indicated that this was not correct in the
pressure range accessible to experiment. A renormalization group
analysis of the problem by Maris \cite{Maris91a} yielded a scaling
relation of $c^{3}\sim P-P_{sl}$ for pressures near $P_{sl}$, which
proved to be very representative of the data of both $^{4}$He and
$^{3}$He. There is some disagreement with this interpretation
\cite{Boronat94b,Solis92}. Accordingly, the $c^{4}$ scaling should
occur only within a short distance from $P_{sl}$ while the $c^{3}$
scaling would be recovered at experimentally observed values of
pressure. However, this seems counter to the renormalization approach
where it would be expected that the exponent should be 4 away from the
critical pressure and 3 some small distance from $P_{sl}$.  At issue
is whether or not the second derivative of pressure with respect to
density is identically zero at the spinodal density. Recent
microscopic calculations by Campbell {\it et al}.\cite{Campbell96} show a
different power law when very near the spinodal point. While the above
discussion is limited to $T=0$~K, the $c^{3}$ scaling behavior is
observed \cite{Hall97} for all temperatures up to the critical
temperature of $T=5.2$~K.

In Fig.~\ref{c3} the isothermal sound velocity from PIMC simulation is
plotted as a function of pressure, being determined numerically from
from the PIMC pressure data,
\begin{equation}
m_{4} c^{2}=\frac{\partial P}{\partial \rho}. \label{speed}
\end{equation}
As can be seen in the figure, the PIMC
data is in agreement with the relation $c^{3}\sim P-P_{sl}$ although
there is some numerical error in taking the derivative. The
experimental data exhibits a smaller slope than the simulation data
but the spinodal pressures (ordinate intercept) are in agreement.


\subsection{Excitation Spectrum}

\begin{figure}
\begin{center}
\epsfig{file=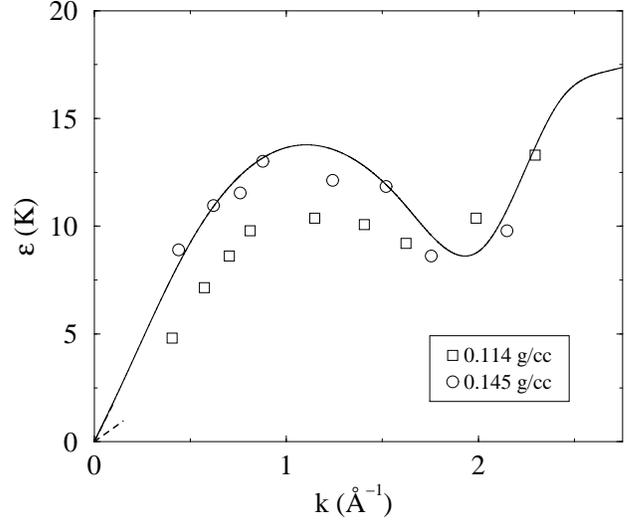,width=3.2in}
\end{center}
\begin{minipage}{3.4in}
\vspace{0.5cm}
\caption{
The excitation spectra from PIMC simulations at a temperature
of 1.0~K and densities of 0.114~g/cc and 0.145~g/cc are compared to the
experimental curve \protect\cite{Donnelly81} (solid line). At a
density of 0.114 g/cc, the system is nearly at the spinodal
pressure. Tangents at $k=0$ are shown (dashed lines), where the
velocity is determined by from Eq.~\ref{speed}. The lowest value of
$k$ for the simulations is determined by the size of the computational
cell.
\label{spectrum}}
\end{minipage}
\end{figure}

The isothermal speed of sound is seen to vanish on the spinodal
line. According to Maris \cite{Hall97}, in the quasi-particle picture
of liquid helium, a transformation of the phonon branch to a free
particle-like dispersion is expected to occur as the liquid spinodal
line is neared. Additionally, at negative pressures, a reduction of
the maxon peak may occur in the excitation spectrum. The effect of
lowering the maxon peak is a spinodal pressure which has a temperature
dependence similar to that in Fig.~\ref{spinodal2}.

At a temperature of 1.0~K the excitation spectrum at two densities has
been calculated using a maximum entropy analysis\cite{Boninsegni96} on
the PIMC intermediate scattering function data and is shown in
Fig.~\ref{spectrum}. Estimates for the uncertainties of these values
are difficult to make. The dynamic structure function is the Fourier
transform of the intermediate scattering function, which is
analytically continued to imaginary time.  In this instance, the
transform is numerically underdetermined with no unique solution. The
peaks in the calculated dynamic structure function reproduce the
experimental spectrum but the calculated widths
of the dynamic structure function are too large in the superfluid
phase \cite{Boninsegni96}. The neutron scattering data at $T<0.35 $~K
and density 0.145~g/cc, compiled by Donnelly \cite{Donnelly81}, is
shown for comparison. The PIMC data at a density of 0.145~g/cc and
pressure $\sim -0.603$~bar agrees with the neutron data.  At a density
of 0.112~g/cc and pressure $\sim -7.72$~bar (very nearly the liquid
spinodal density), there is an obvious decrease in the maxon peak but
since the maxon peak is density dependent, it is not clear whether the
effect is solely the result of the proximity to the spinodal. The
phonon part of the spectrum is at momenta less then the lowest wave
vector present in the system being simulated. However, the slope of
the spectrum at zero wavevector is known, Eq.~\ref{speed}, and is
shown (dashed lines) for each value of the density.


\section{Conclusions}

The spinodal line, the superfluid transition and the excitation
spectrum for liquid helium at negative pressures have been
investigated using PIMC.  The spinodal line at low temperatures shows
substantial agreement with the density functional calculations of
Ref.~\onlinecite{Guirao92} as well as at temperatures above the
superfluid transition. The calculated temperature independence at low
temperatures and the sudden increase in pressure as the superfluid
transition is neared is consistent with the conjectures of
Ref.~\onlinecite{Hall97}. At higher temperatures there is general good
agreement of the spinodal pressure with all calculations in the normal
fluid phase. The superfluid transition is found to be consistent with
the experimental transition values and approaches the spinodal line
near $2.2$ K. Finally, the excitation spectrum exhibits changes in the
maxon area consistent with the low temperature behavior of the liquid
spinodal line.


\acknowledgments

The authors would like to acknowledge Manuel Baranco for providing
their density functional code, and Humphrey Maris for useful
discussions during the course of this work. This research was
supported by NASA's Microgravity Research Fundamental Physics
Program, NAG3-1926.  The simulations were performed at the
National Computational Science Alliance (NCSA).


\begin{references}

\bibitem{Maris93} H.J. Maris, S. Balibar and M.S. Pettersen, J. Low Temp. Phys.  {\bf 93}, 1069 (1993).
\bibitem{Maris96} H.J. Maris, Czech. J. Phys. {\bf 46}, 2943 (1996).
\bibitem{Balibar95} S. Balibar, C. Guthmann, H. Lamba\'{r}e, P. Roche, E. Rolley and H.J. Maris, J. Low Temp. Phys. {\bf 101}, 271 (1995).
\bibitem{Balibar98} S. Balibar, F. Caupin, P. Roche and H.J. Maris, J. Low Temp. Phys. {\bf 113}, 459 (1998).
\bibitem{Maris94a} H.J. Maris, J. Low Temp. Phys. {\bf 94}, 125 (1994).
\bibitem{Hall95} S.C. Hall, J. Classen, C.-K. Su and H.J. Maris, J. Low Temp.  Phys. {\bf 101}, 793 (1995).
\bibitem{Classen98b} J. Classen, C.K. Su, M. Mohazzab and H.J. Maris, Phys. Rev. B {\bf 57}, 3000 (1998).
\bibitem{Classen98a} J. Classen, C.K. Su, M. Mohazzab and H.J. Maris,
 J. Low Temp. Phys. {\bf 110}, 431 (1998).
\bibitem{Lambare98} H. Lamba\'{r}e, P. Roche, S. Balibar, H.J. Maris, O.A. Andreeva, C. Guthmann, K.O. Keshishev and E. Rolley, Eur. Phys. J. B {\bf 2}, 391 (1998).
\bibitem{Pettersen94a} M.S. Pettersen, S. Balibar and H.J. Maris, Phys. Rev.  B {\bf 49}, 12062 (1994).
\bibitem{Guirao92} A. Guirao, M. Centelles, M. Barranco, M. Pi, A. Polls and X. Vi\~{n}as, J. Phys.: Condens. Matter {\bf 4}, 667 (1992).
\bibitem{Hall97} S.C. Hall and H.J. Maris, J. Low Temp. Phys. {\bf 107}, 263  (1997).
\bibitem{Boronat94a} J. Boronat and J. Casulleras, Phys. Rev. B {\bf 49}, 8920 (1994).
\bibitem{Boronat94b} J. Boronat, J. Casulleras and J. Navarro, Phys. Rev. B {\bf 50}, 3427 (1994).
\bibitem{Solis92} M.A. Sol\'{i}s and J. Navarro, Phys. Rev. B {\bf 45}, 13081 (1992).
\bibitem{Dalfovo95} F. Dalfovo, A. Lastri, L. Pricaupenko, S. Stringari and J. Treiner, Phys. Rev. B {\bf 52}, 1193 (1995).
\bibitem{Campbell96} C.E. Campbell, R. Folk and E. Krotscheck, J. Low Temp. Phys. {\bf 105}, 13 (1996).
\bibitem{Ceperley95} D.M. Ceperley, Rev. Mod. Phys. {\bf 67}, 279 (1995).
\bibitem{Gordillo98} M.C. Gordillo and D.M. Ceperley, Phys. Rev. B {\bf 58}, 6447 (1998).
\bibitem{Aziz95} R. A. Aziz, A. R. Janzen, M. R. Moldover, Phys. Rev. Letts. {\bf74},1586 (1995).
\bibitem{Ceperley89} D.M. Ceperley and E.L. Pollock, Phys. Rev. B {\bf 39}, 2084 (1989).
\bibitem{Pollock92} E.L. Pollock and K.J. Runge, Phys. Rev. B {\bf 46}, 3535 (1992).
\bibitem{Pollock87} E.L. Pollock and D.M. Ceperley, Phys. Rev. B {\bf 36}, 8343 (1987).
\bibitem{Cha91} Min-Chul Cha {\it et al}, Phys. Rev. B {\bf 44}, 6883 (1991).
\bibitem{Gruter97} P. Gr\"{u}ter, D. M. Ceperley and F. Lalo\"{e}, Phys. Rev. Lett. {\bf 79}, 3549 (1997).

\bibitem{Maris89} H.J. Maris and Q. Xiong, Phys. Rev. Lett. {\bf 63}, 1078 (1989).
\bibitem{Maris91a} H.J. Maris, Phys. Rev. Lett. {\bf 66}, 45 (1991).
\bibitem{Maris91b} H.J. Maris, in {\it Excitations in Two-Dimensional and Three-Dimensional Quantum Fluids} edited by A.G.F. Wyatt and H.J. Lauter (1991), p. 107.
\bibitem{Donnelly81} R.J. Donnelly, J.A. Donnelly, and R.N. Hills,
J. Low Temp. Phys. {\bf 44}, 471 (1981).
\bibitem{Boninsegni96} M. Boninsegni and D.M. Ceperley, J. Low Temp. Phys.  {\bf 104}, 339 (1996).

\end{references}
\end{document}